\documentclass[aps,prl,reprint,superscriptaddress]{revtex4-1}

\usepackage{tikz}
\usetikzlibrary{arrows,snakes,backgrounds}
\usepackage{amsmath}
\usepackage{amscd}
\usepackage{amsfonts}

\newcommand{\ket}[1]{\rvert #1\rangle }
\newcommand{\up}{\mid\uparrow\rangle}
\newcommand{\down}{\mid\downarrow\rangle}

\newcommand{\expectation}[1]{\langle #1 \rangle }
\newcommand{\abs}[1]{\left\lvert #1 \right\rvert}
\newcommand{\Sr}{$^{88}Sr^+$}
\newcommand{\Hz}{\ Hz}

\begin{document}


\title{Single-Spin Spectrum-Analyzer for a Strongly Coupled Environment}


\author{Shlomi Kotler}
\email{shlomi.kotler@weizmann.ac.il}
\homepage{www.weizmann.ac.il/complex/ozeri}
\affiliation{Department of Physics of Complex Systems, Weizmann Institute of Science, Rehovot 76100, Israel}

\author{Nitzan Akerman}
\affiliation{Department of Physics of Complex Systems, Weizmann Institute of Science, Rehovot 76100, Israel}
\author{Yinnon Glickman}
\affiliation{Department of Physics of Complex Systems, Weizmann Institute of Science, Rehovot 76100, Israel}
\author{Roee Ozeri}
\affiliation{Department of Physics of Complex Systems, Weizmann Institute of Science, Rehovot 76100, Israel}

\date{\today}

\begin{abstract}
A qubit can be used as a sensitive spectrum analyzer of its environment. Here we show how the problem of spectral analysis of noise induced by a strongly coupled environment can be solved for discrete spectra. Our analytical model shows non-linear signal dependence on noise power, as well as possible frequency mixing, both are inherent to quantum evolution. This model enabled us to use a single trapped ion as a sensitive probe for strong, non-Gaussian, discrete magnetic field noise. To overcome ambiguities arising from the non-linear character of strong noise, we develop a three step noise characterization scheme: peak identification, magnitude identification and fine-tuning. Finally, we compare experimentally equidistant versus Uhrig pulse schemes for spectral analysis. The method is readily available to any quantum probe which can be coherently manipulated.
\end{abstract}


\maketitle

The ability of a quantum system to withstand noise is characterized by its decoherence rate; the rate at which superpositions deteriorate. Hahn's discovery~\cite{Hahn1950} of the echo technique showed that reducing decoherence can be achieved by external modulation, e.g in the case of spins, performing a single spin flip during the experiment. Since then, the idea of using external modulation to prolong coherence, known as Dynamic Decoupling~\cite{Viola1998,Ban1998,Viola1999,Tombesi1999}, has been well developed to include many pulses~\cite{CarrPurcell1954,GillMeiboom1958} and different modulation schemes~\cite{Lidar2005,uhrig2007,gordon2008,Uys2009,jens2010}.

These techniques rely on the condition that a quantum system, modulated at frequency $f$, is most influenced by the noise power spectral density at $f$. For a two-level system experiencing phase noise which is either Gaussian or weak (perturbative), this condition takes the following integral overlap form~\cite{KofmanKurizki2001,KofmanKurizki2004,goren:2007}:
\begin{equation} \label{eq:pert}
R(T)\equiv -\frac{1}{T}\ln \rho_{12}=\int_{-\infty}^\infty df S(f) \abs{F_T(f)}^2
\end{equation}
where $\rho$ is the system density matrix, $R(T)$ the decoherence rate, $S(f)$ the noise power spectral density and $F_T(f)$ is the Fourier transform of the applied modulation for an experiment of length $T$. By generating modulation schemes which have a small overlap with the noise spectrum, $R$ diminishes. Therefore, different dynamical decoupling schemes are optimal for different noise spectra. For a detailed account, see \cite{Biercuk2011} and references therein.

Can one use decoherence as a measurement tool? For that matter, Eq. \eqref{eq:pert} can be considered in the inverse manner. Instead of prolonging coherence by separating the modulation and noise spectra one can focus $F_T(f)$ about a given spectral component at say, $f_0$, thereby extracting $S(f_0)$. This idea of a qubit-based spectrum analyzer has been suggested in the context of different qubit technologies~\cite{Lasic2006,dasSarma2008,deSousa2009,Hollenberg2009} and has been recently analyzed for different types of noise spectra~\cite{Yuge2011}.

Experimental realizations of spectral analysis through spin decoherence spectroscopy were performed with different technologies; for example, in trapped ions~\cite{Biercuk2009}, cold atomic ensembles~\cite{SagiPT2010,Almog2011}, nitrogen-vacancy centers in diamonds~\cite{deLange2010}, super-conducting flux qubits~\cite{Bylander2011} and NMR experiments in molecules~\cite{Alvarez2011}.  The decoherence spectrum can be used to fix parameters in a known noise model, or reconstruct it through the inversion of the decoherence-spectrum relation~\cite{Alvarez2011,Almog2011}.

The operation of such qubit spectrum analyzers relies on the validity of Eq. \eqref{eq:pert}, i.e. a linear response in the spectrum as in the case of weak or Gaussian noise. Otherwise, the qubit evolution can be significantly non-linear in Hamiltonian terms. In which case, the relation between measured decoherence and noise is very hard to calculate.

It turns out, as will be shown in this paper, that spectral analysis in the strong noise limit can be significantly simplified for discrete spectra. This is reminiscent of the use of simple frequency analysis tools which enabled Babylonian astronomers to accurately predict the timings of lunar and solar eclipses~\cite{Neugebauer1969} and 19$^\textrm{th}$ century scholars to provide tide predictions for various coasts and harbors~\cite{Parker2011}. This is despite the fact that nonlinear evolution is present in planet and ocean dynamics as well.

The merit of using a strongly coupled qubit can be understood in terms of the trade-off between signal sensitivity and spectral resolution using the Cram\'er-Rao bound. For an experiment time $T$ and noise amplitude $N$ the weak assumption is equivalent to a small noise index $\eta\equiv NT\ll 2\pi$. If coherence is estimated by a quantum projective measurement, the optimal amplitude signal-to-(projection)noise ratio of a continuous spectrum is obtained when $\eta\to 0$, consistent with the weak assumption. For large $N$, however, this will impose a short experiment time thereby limiting the spectral resolution which scales as $1/T$. Better spectral resolution requires departure from the weak limit.

For the case of discrete spectra, the Cram\'er-Rao bound implies that both the amplitude signal-to-noise ratio and the spectral resolution are optimal for $\eta\ge 2\pi$. Moreover, the spectral resolution attains an enhancement factor and scales as $1/(T\sqrt{\eta})$ due to the non-linear response of the coherence with respect to noise amplitude. Spectral analysis of discrete spectra should therefore benefit from operating in the strong noise limit.

In this work we describe a simple analytical model extending Eq. \eqref{eq:pert} to the non-perturbative, non-Gaussian discrete case. We further show how one can use this theory to identify the noise spectral components and measure their magnitude in typical noise scenarios. We apply our spectral analysis scheme, using a single-trapped ion, to analyze the discrete spectrum of magnetic field noise in our lab.

We focus on a two level quantum probe described by $\ket{\psi(t)}=\alpha\up+e^{i\phi}\beta\down$ and governed by a Hamiltonian $H=\hbar(N(t)\hat{\sigma}_z+\Omega(t)\hat{\sigma}_x)/2$ where $N(t)$ is classical dephasing noise and $\Omega(t)$ is the spectrum analyzer modulation. We assume no spin relaxation processes. Our purpose is to use the modulation $\Omega(t)$ to quantify the noise $N(t)$.

For a probe initialized to $\ket{\psi_0}=(\up+\down)/\sqrt{2}$ the superposition relative phase at time $T$ is~\cite{Kotler2011}:
\begin{equation}\label{eq:lockin}
\phi(T)=\int_0^T dt N(t) F(t)=\int_{-\infty}^{\infty} df N(f)F_{T}(f)
\end{equation}
where $F(t)\equiv\cos(\int_0^tdt' \Omega(t'))$, $N(f),F_T(f)$ are the respective Fourier transforms, the latter calculated on a truncated experiment window of length $T$. 

Phase coherence is obtained by averaging over noise realizations, $A\equiv\expectation{e^{i\phi}}$. The assumption that noise is either weak or Gaussian translates into $\expectation{e^{i\phi}}=e^{-\expectation{\phi^2}/2}$. Here, the decoherence rate $R$ is proportional to $\expectation{\phi^2}$, the variance of the superposition phase imposed by noise. Combined with Eq. \eqref{eq:lockin}, relation \eqref{eq:pert} follows.

To calculate the phase coherence without assuming that noise is Gaussian or perturbative, we assume discreteness: $N(t)=\sum_{k=1}^d \abs{N_k}\cos(2\pi f_kt +\alpha_k)$. For a single noise component, according to \eqref{eq:lockin}, $\phi=\abs{N_0F_T(f_0)}\cos(\alpha)$ so $\expectation{e^{i\phi}}=(2\pi)^{-1}\int_0^{2\pi}d\alpha e^{i\phi}=J_0(\abs{N_0F_T(f_0)})$ where $J_0$ is the zeroth Bessel function of the first kind. For an ideal sinusoidal modulation at $f_{0}$, $A=J_0(\abs{N_{k_0}}T)$. The weak limit is valid only when $J_0$ can be well approximated to second order in its argument. Moreover, when the noise index $\eta\equiv \abs{N_{k_0}}T$ crosses $z_0\approx 2.4$, the first zero of $J_0(x)$, coherence becomes negative, i.e. the phase superposition partially refocuses close to $\pi$. Notice that in general $\eta= c\abs{N_{k_0}}T$ where c is a numerical constant depending on the modulation shape. For the square wave modulation $c=2/\pi$ and for Uhrig modulation $c\approx 0.42$. Such single Bessel behavior due to a single mechanical resonance of a cantilever coupled to a nitrogen-vacancy (NV) center was recently observed~\cite{Kolkowitz2012}.

In the case of more than one noise component, the coherence behavior takes a product form over all noise components. Assuming $\alpha_k\in[0,2\pi]$ are uniform mutually independent variables,
\begin{equation}\label{eq:bessel}
A(T)\equiv\expectation{e^{i\phi}}= \prod_k J_0\left(\abs{N_k F_T(f_k)}\right)
\end{equation}
This equation coincides with Eq. \eqref{eq:pert} by Taylor expanding the Bessel functions to second order and recalling that $R\propto -\ln A$. Equation \eqref{eq:bessel} is the main tool of our noise spectral estimation method.

The strong noise limit also reveals frequency mixing if we allow correlations in the $\alpha_k$-s. Whenever an integer combination of the noise frequencies is nulled $\sum_k h_k f_k=0$, additional Bessel product terms affect the coherence,
\begin{equation}\label{eq:genbessel}
A(T)=\sum_{\substack{h_1,\ldots,h_d\\ \Sigma h_kf_k =0\\ \Sigma h_k\textrm{ even}}}(-1)^{\frac{1}{2}\Sigma h_k}\cos(\Sigma h_k\alpha_k)\prod_{k=1}^d J_{h_k}(\abs{N_k F_T(f_k)})
\end{equation}
where the $h_k$ are integers and $J_{h_k}$ the corresponding Bessel functions of first kind. The dominant summand corresponding to $h_1=\ldots=h_d=0$ coincides with Eq. \eqref{eq:bessel}. By focusing the modulation at a single frequency, as in our experiment, all the higher Bessel terms can be neglected, and information on the phase relation between different spectral components is lost. One will be able to retrieve it via the $\cos(\Sigma h_k \alpha_k)$ term by using a multi-tonal modulation.

\begin{figure}[!hbtp]
 \includegraphics{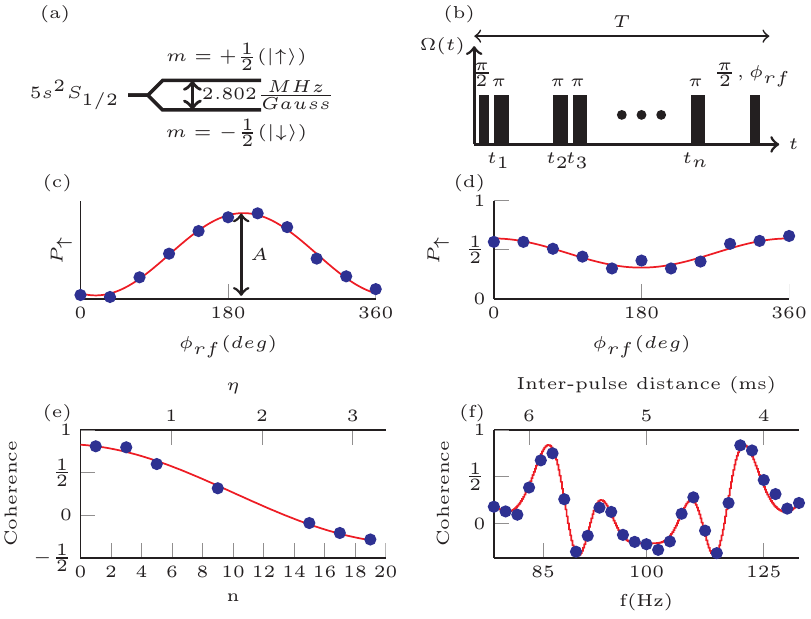}
 \caption{\label{fig:experiment} (a) \Sr ground state manifold comprising the Zeeman sensitive quantum probe. (b) Typical experimental sequence with $n$ modulation pulses sandwiched between two $\pi/2$ pulses with a relative $\phi_{rf}$ phase. (c,d) Probability of finding the probe in the $\up$ state at the end of a sequence vs. the rf relative phase. Red line is a best fit to $P_{\uparrow}=1/2-A/2\cos(\phi_{rf})$, $A$ is the coherence. Plot 1c was measured with $n=1$ pulses and 1d with $n=19$ equidistant pulses with inter pulse distance of $5\ ms$. (e) Coherence vs. number of equidistant pulses $n$. The first and last points correspond to the fringes in 1c and 1d. Red line is a single parameter best-fit to Eq. \eqref{eq:bessel}. Inversion of the fringe contrast in d correspond to negative coherence. (f) Coherence vs. modulation frequenct, with $n=11$ equidistant pulses. Although the spectrum shows five spectral features, it corresponds to a single, highly non-perturbative noise component. Red line is a single parameter best-fit to Eq. \eqref{eq:bessel}.}
\end{figure}

Our system is comprised of the two spin states of the electronic ground level of a single $^{88}Sr^+$ ion, $\up=\ket{5s_{1/2}, J=1/2, M_J=1/2}$ and $\down=\ket{5s_{1/2}, J=1/2, M_J=-1/2}$. This Zeeman sub-manifold is first order sensitive to external magnetic fields. The dominant noise we measured was magnetic field fluctuations $B(t)$ due to power line harmonics rendering a discrete noise spectrum, $N(t)=g\mu_B B(t)/\hbar$, where $g$ is the Land\'e g-factor, $\mu_B$ the Bohr magneton and $\hbar$ the Planck constant divided by $2\pi$ (see Fig. \ref{fig:experiment}a). We performed spin rotations by pulsing a resonant rf magnetic field. Rotation angles and rotation axes were controlled by tuning the pulse duration and the rf field phase, $\phi_{rf}$, respectivley. State initialization and measurement were performed by optical pumping and state-selective fluorescence correspondingly~\cite{Akerman2011,Keselman2011}.

To measure the phase coherence we performed a Ramsey-type experiment as shown in Fig. \ref{fig:experiment}b. A modulation $\Omega(t)$ of length $T$ is sandwiched between two $\pi/2$ pulses, differing by a relative phase $\phi_{rf}$. We then measured the probability of the ion to be in the $\up$ state $P_{\uparrow}$ as a function of $\phi_{rf}$. A fit to $P_{\uparrow}=\frac{1}{2}-\frac{A}{2}\cos\phi_{rf}$ yields an experimental estimate of the phase coherence $A$. Examples of such fringes are shown in Fig. \ref{fig:experiment}c and \ref{fig:experiment}d. In all cases, $\Omega(t)$ was a train of $\pi$ pulses at different times and possibly different rotation axes.

A first distinctive characteristic of non-perturbativity is the negative values of the coherence (noise index $\eta>z_0$), shown in \ref{fig:experiment}e. Here we fixed the modulation frequency at $f_{mod}=100Hz$ while increasing $n$, the number of equidistant pulses. As seen, the fringe contrast with $n=19$ (shown in Fig. \ref{fig:experiment}d) is inverted with respect to $n=1$ (shown in Fig. \ref{fig:experiment}c). A fit to Eq. \eqref{eq:bessel} is shown by the red line, assuming a single spectral component at $100\ Hz$, with $N_0$ as a single fit parameter and results in $B_{100 Hz}=3.0 (2)\mu G$.

A second mark of non-perturbativity is that multiple spectral features can arise from a single noise components, as shown in Fig. \ref{fig:experiment}f. The number of pulses is fixed at $n=11$ and the modulation frequency is scanned around $f=100\ Hz$. The spectrum shows a "power broadened" spectral feature around $100\ Hz$ with five coherence minima. Unlike the perturbative case these do not correspond to five different spectral components but rather to a broadened response to a magnetic field monotone. Again Eq. \eqref{eq:bessel} with $N_0$ as a single fit parameter shown by the red line is used, yielding $B_{100\ Hz}=15.3(3)\mu G$. This noise amplitude corresponds to a noise index of $\eta=10.3(2)$, well in the strong noise regime. The noise amplitudes extracted from the data shown \ref{fig:experiment}e and \ref{fig:experiment}f are very different as these data sets were taken at different times.

\begin{figure}[!hbtp]
 \includegraphics{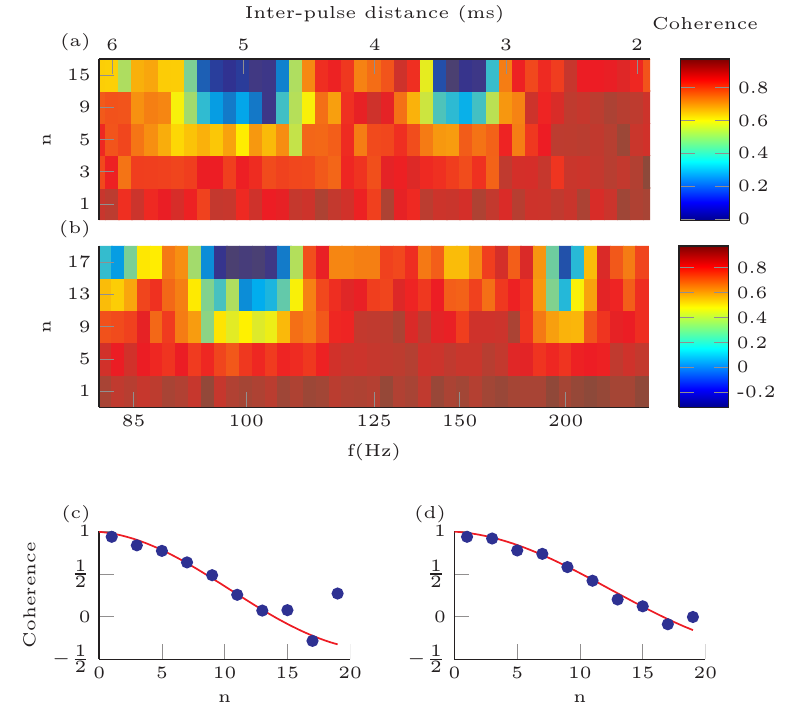}
 \caption{\label{fig:elements} Spectral peak identification. (a) Two scans taken several months apart. Each scan is a colormap of coherence vs. the number of pulses $n$ and the modulation frequency. The upper scan shows clear features at $100\ Hz$ and $150\ Hz$. The latter nearly vanishes in the lower scan and a new $200\ Hz$ component appears. (b) Magnitude extraction of a single noise component. The number of pulses $n$ is varied while the modulation frequency is kept fixed at $100\ Hz$ until a zero-crossing is reached. From this zero crossing the field magnitude can be estimated. Red line is a fit to Eq. \eqref{eq:bessel} with a single fit parameter (c) Same as b for the $200\ Hz$ component.}
\end{figure}

To practically estimate a multi-tone discrete spectrum we first identify the frequencies $f_k$ of its components. In any modulation scheme, the peak of the modulation $F_T(f)$ increases linearly with the total experiment time $T$ while improving spectral resolution. To identify the different noise components we therefore modulated the probe at different frequencies. For each modulation frequency the number of pulses was increased until the different noise components emerged. Examples are shown in figures \ref{fig:elements}a and \ref{fig:elements}b.
Here, the measured coherence is shown vs. the (equidistant) modulation frequency and the number of $\pi$ pulses. As the number of pulses was increased clear spectral features emerged. The two data sets were measured four months apart with a different magnetic environment; the spectral response at $150\ Hz$ which is clear in Fig. \ref{fig:elements}a almost vanishes in \ref{fig:elements}b where a new $200\ Hz$ component appeared.

Once the component frequencies $\{ f_k\}$ have been determined, the multiplicative structure of equation \eqref{eq:bessel} is used to determine their magnitudes $N_k$. Whenever the coherence $A(T)$ crosses zero, with high certainty, only one of the Bessel functions in the product is nulled. If the modulation is centered about $f_k$, increasing the experiment time $T$ until the first zero crossing occurs implies that the corresponding Bessel has been nulled and provides an estimate for $N_k$. We used this method to extract the magnitudes of the $100\ Hz$ and $200\ Hz$ components identified in Fig. \ref{fig:elements}b. We focused an equidistant modulation at $f=100Hz$ ($200Hz$) while increasing the number of pulses, $n$. Coherence vs. $n$ is shown in Figure \ref{fig:elements}c (\ref{fig:elements}d). From the zero crossing we estimate a noise amplitude of $2.6(2)\ \mu G$ ($5.0(2)\ \mu G$). This is in reasonable agreement with a best-fit to Eq. \eqref{eq:bessel} shown by the red line, yielding $2.9 (1) \mu G$ ($4.9(3)\ \mu G$).

\begin{figure}[!htbp]
\includegraphics{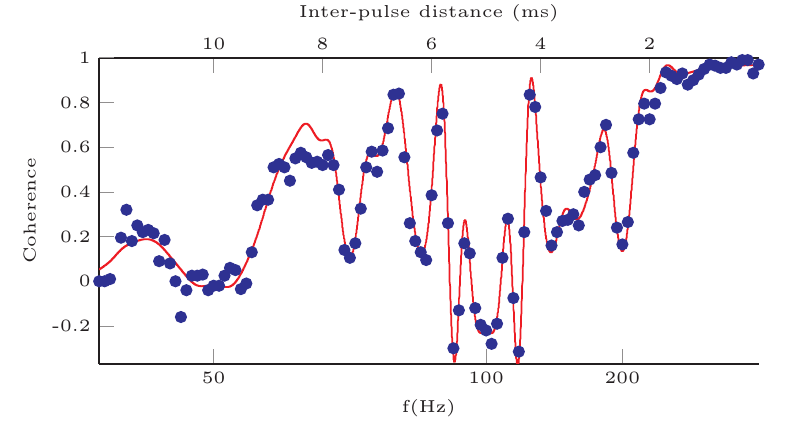}
\caption{\label{fig:fits}Fine tuning of noise magnitudes. Once noise components have been identified in frequency and magnitude a fine tuning estimate is obtained by scanning the modulation frequency and fitting to the full spectrum to Eq. \eqref{eq:bessel} (red line). Here we use $n=11$ equidistant pulse sequence.}
\end{figure}

The last stage of spectral characterization is fine tuning of the estimated noise magnitudes with a full fit procedure, using the previously estimated field magnitudes as a starting point. Such a fit to Eq. \eqref{eq:bessel} is shown in figure \ref{fig:fits} with five fit parameters $B_{50 Hz}=2.0(1)\mu G$, $B_{100 Hz}=15.4(4)\mu G$, $B_{150 Hz}=4.2(3)\mu G$, $B_{200 Hz}=6.3(3)\mu G$ and a slowly varying field, $(g\mu_B B_{slow}/h)f_{slow}=66(2)Hz^2$. The non-perturbative nature of the spectrum is quantified by the corresponding noise indices: $\eta_{50}=2.7(1)$, $\eta_{100}=10.4(3)$, $\eta_{150}=1.9(1)$, $\eta_{200}=2.1(1)$.

\begin{figure*}[!htbp]
\includegraphics{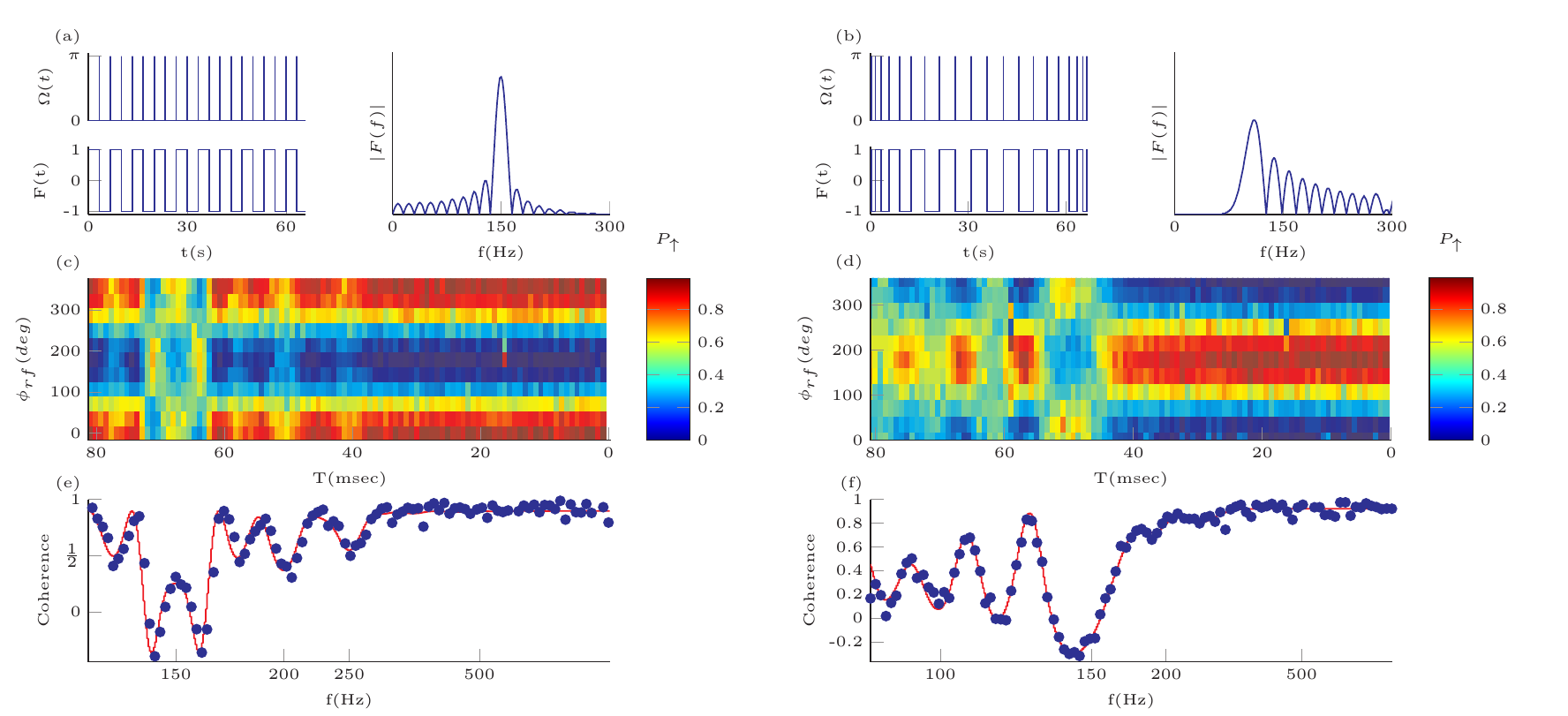}
\caption{\label{fig:urig}Comparison of two modulation schemes: equidistant and Uhrig. (a) The spectrum of equidistant modulation centered at $150\ Hz$ with $19$ pulses. (b) Same as a, for the $20$ pulses Uhrig scheme with the same total duration of $66.7\ ms$. The modulation spectrum has much more energy in high frequencies, and its peak is shifted to $109.6Hz$. (c) Each column is a scan of $\phi_{rf}$ at $n=19$ equidistant pulses and a fixed $T$. The total experiment time $T$ is varied from column to column and the corresponding modulation frequency is shown in e. (d) Same as c, for an Uhrig scheme of $20$ pulses. (e) Each point corresponds to the fringe contrast of corresponding column in c. Clear peaks are pronounced in $150\Hz$, $200\Hz$ and $250\Hz$. Red line is a fit to Eq. \eqref{eq:bessel}. (f) Same as e, for the Uhrig scheme. As the modulation mixes neighboring frequencies, it is harder to identify spectral peaks compared to the equidistant case. The fit indicated by the red line yields field magnitudes consistent with the equidistant modulation fit in e.}
\end{figure*}

What modulation suits best for spectrum estimation? Ideally, one would require a Fourier-limited sinc around the modulation frequency, as used in~\cite{Almog2011}. It is, however, more convenient experimentally to use a stream of $\pi$ pulses. For the purpose of noise spectroscopy Yuge et. al.~\cite{Yuge2011} suggested an equidistant pulse scheme while Cywi\ifmmode \acute{n}\else \'{n}\fi{}ski et. al.~\cite{dasSarma2008} suggested the Uhrig~\cite{uhrig2007} scheme. Previously, these schemes have been compared experimentally with trapped ions~\cite{Biercuk2009} and solid state NMR~\cite{Alvarez2011PRA} in the context of optimized dynamical decoupling.  Here we use both schemes for the purpose of spectral estimation. A comparison of typical modulation spectra of the two is shown in Fig. \ref{fig:urig}a and \ref{fig:urig}b; in both cases $19$ pulses are used for a total experiment time of $T=66.7\ ms$.

In terms of spectral filters, the equidistant case is more intuitive since its peak is centered around $f_{mod}=(n+1)/2T$ while in the Uhrig case, the spectral peak is shifted towards lower frequencies. Moreover its spectral content at high frequencies is much greater than the equidistant case. However at low frequencies, Uhrig modulation decays faster. It appears that Uhrig modulation should be more suited when one wants to measure discrete high frequency components in the presence of unwanted low-frequency noise. Otherwise equidistant modulation is simpler to use and interpret.

The interpretational simplicity of equidistant modulation is pronounced in the two-dimensional scan shown in Fig. \ref{fig:urig}c. For each fixed experiment duration, $T$, a phase scan is shown as a column in figure \ref{fig:urig}c. The contrast of each column is obtained from a fit procedure and displayed as a single data point in Fig. \ref{fig:urig}e correspondingly. The noise spectral peaks can be easily identified at $150\ Hz$, $200\ Hz$ and $250\ Hz$. The spectral shape of the Uhrig modulation (Fig. \ref{fig:urig}b) mixes nearby frequency components of the noise, as seen in figure \ref{fig:urig}d and \ref{fig:urig}f.

Quantitatively, with our noise profile, both modulation schemes performed equally well and show agreement in the extracted spectrum (figure \ref{fig:urig}e,f). This is summarized in the table \ref{tbl}.
\begin{table}
\caption{\label{tbl} Fine tuning of field magnitudes using equidistant and Uhrig modulation.}
\begin{ruledtabular}
\begin{tabular}{cccc}
$f_{noise}(Hz)$ & equidist. tune($\mu G$)& Uhrig tune($\mu G$)& $\eta$\\
\hline
100& -      & 2.7(2)&  1.3(1) \\ 
150& 9.0(3) & 8.7(2)&  6.7(2) \\
200& 2.8(2) & 2.8(5)&  1.6(1) \\
250& 3.0(3) & 3.3(6)&  1.3(1) \\
\end{tabular}
\end{ruledtabular}
\end{table}
The Uhrig scheme reveals information on an additional frequency component, $B_{100Hz}$, due to the shift of its spectral peak to lower frequency noise as compared with the equidistant case.

In conclusion, we have developed and used an analytical model for spectral noise estimation of non-perturbative, non-Gaussian, discrete dephasing noise. We used our scheme to extract the magnetic field noise spectral components at the power line harmonics in various real lab scenarios. In fact, this enabled us to calibrate our magnetic field compensation system to reduce noise components to the $\mu G$ level, thereby reaching $1.4\ s$ of coherence with a Zeeman sensitive qubit~\cite{Kotler2011}. We expect this model to be useful for other discrete noise scenarios. One example is spontaneous $\alpha$-oscillations in brain activity measured using magnetoencephalography (MEG)~\cite{Sander2012}. Another example is the study of decoherence of a single NV center induced by a finite number of $^{13}$C nuclear spins in diamond~\cite{Wrachtrup2012} or the discrete mechanical resonances of a cantilever coupled to an NV center~\cite{Kolkowitz2012}.

\begin{acknowledgments}
We gratefully acknowledge the support by the Israeli Science foundation, the Minerva foundation, the German-Israeli foundation for scientific research, the US-Israel Binational Science Foundation, the Crown photonics center and David Dickstein, France.
\end{acknowledgments}

\end{document}